\documentclass{tcibook}
\usepackage{fancyhea}
\usepackage{work}
\usepackage{bm}       
\usepackage{graphicx}
\usepackage{hyperref}      

\newcommand{\nc}{\newcommand}  

\def\beq{\begin{equation}}
\def\eeq#1{\label{#1}\end{equation}}
\def\eeqn{\end{equation}}

\newenvironment{Eqnarray}%
   {\arraycolsep 0.14em\begin{eqnarray}}{\end{eqnarray}}
\def\beqa{\begin{Eqnarray}}
\def\eeqa#1{\label{#1}\end{Eqnarray}}
\def\eeqan{\end{Eqnarray}}

\nc{\ra}{\rightarrow}  
\nc{\slsh}{\slash\hspace*{-0.22cm}}
\def\Re{{\cal R \mskip-4mu \lower.1ex \hbox{\it e}\,}}
\def\Im{{\cal I \mskip-5mu \lower.1ex \hbox{\it m}\,}}

\nc{\vev}[1]{ \left\langle {#1} \right\rangle }
\nc{\bra}[1]{ \langle {#1} | }
\nc{\ket}[1]{ | {#1} \rangle }
\nc{\fb}{\,{\rm fb}^{-1}}
\nc{\ev}{{\rm eV}}
\nc{\kev}{{\rm keV}}
\nc{\Mev}{{\rm MeV}}
\nc{\gev}{{\rm GeV}}
\nc{\tev}{{\rm TeV}}
\nc{\mev}{{\rm MeV}}

\def\del{\partial}
\def\Dslash{\not{\hbox{\kern-4pt $D$}}}
\def\dslash{\not{\hbox{\kern-2pt $\del$}}}
\def\pslash{\not{\hbox{\kern-2pt $p$}}}
\def\ETmiss{ \not{\hbox{\kern-4pt $E$}}_T }

\def\msb{{\bar{\ssstyle M \kern -1pt S}}}

\begin{document}

\def\bibname{References}
\bibliographystyle{plain}

\raggedbottom

\pagenumbering{roman}

\parindent=0pt
\parskip=8pt
\setlength{\evensidemargin}{0pt}
\setlength{\oddsidemargin}{0pt}
\setlength{\marginparsep}{0.0in}
\setlength{\marginparwidth}{0.0in}
\marginparpush=0pt


\pagenumbering{arabic}

\renewcommand{\chapname}{chap:intro_}
\renewcommand{\chapterdir}{.}
\renewcommand{\arraystretch}{1.25}
\addtolength{\arraycolsep}{-3pt}




\chapter{Computing Frontier: Software Development, Staffing and Training}
\label{chap:CpFI4}
\def \HEP {HEP }

\begin{center}\begin{boldmath}



\begin{center}
{\large Conveners: David Brown$^1$, Peter Elmer$^2$\\
Observer: Ruth Pordes$^3$\\
Contributing Authors: David Asner$^4$, Gregory Dubois-Felsmann$^5$, V. Daniel Elvira$^3$, Robert Hatcher$^3$, Chris Jones$^3$, Robert Kutschke$^3$, David Lange$^6$, Elizabeth Sexton-Kennedy$^3$, Craig Tull$^1$}\\
\bigskip
$^1${\it Lawrence Berkeley National Laboratory (LBNL)}\\
$^2${\it Department of Physics, Princeton University}\\
$^3${\it Fermi National Accelerator Laboratory (FNAL)}\\
$^4${\it Pacific Northwest National Laboratory (PNNL)}\\
$^5${\it SLAC National Accelerator Laboratory (SLAC)}\\
$^6${\it Lawrence Livermore National Laboratory (LLNL)}\\
\end{center}
\bigskip


\begin{large} {\bf Executive Summary} \end{large}


\end{boldmath}\end{center}

Success of \HEP science will continue to depend critically on computing.
Managing the human activities (software development and management,
training and staffing) is an important part of that.  
Based upon our own experiences, and from
discussions with members of the \HEP community,
we have identified the following main goals for the next decade in the area
of software, staffing and training:

\begin{itemize}
\item Goal: To maximize the scientific productivity of our community
in an era of reduced resources, we must use
software development strategies and staffing models that will result in products
that are generally useful for the wider \HEP community.
\item Goal: We must respond to the evolving technology market, especially
with respect to computer processors, by
developing and evolving software that will perform with optimal efficiency
in future computing systems.
\item Goal: We must insure that our developers and users will have the
training needed to create, maintain, and use the increasingly complex software
environments and computing systems that will be part of future \HEP projects.
\end{itemize}

Some specific recommendations we feel will help achieve these goals are:

\begin{itemize}
    \item Software Management, Toolkits and Reuse
    \begin{itemize}
        \item Continue to support established toolkits (Geant4, ROOT, ...)
        \item Encourage the creation of new toolkits from existing successful common software (generators, tracking, ...)
        \item  Allow flexible funding of software experts to facilitate transfer of software and sharing of technical expertise between projects
        \item Facilitate code sharing through open-source licensing and use of publicly-readable repositories.
        \item Consolidate and standardize software management tools to minimize cross-project ``friction''
    \end{itemize}

    \item Software Development for new Hardware Architectures
    \begin{itemize}
        \item Significant investments in software are needed to adapt to the evolution of computing processors, both as basic R\&D into appropriate
techniques and as re-engineering ``upgrades''
        \item New software should be designed, and existing software re-engineered, to expose parallelism at multiple levels
        \item Develop flexible software architectures that can exploit efficiently a variety of possible future hardware options
    \end{itemize}    

    \item Staffing
      \begin{itemize}
          \item Recognize software efforts as sub-projects of the project
          \item Integrate computing professionals as part of the project team, over the life of the project or collaboration
          \item Integrate software professionals with scientist developers to insure software meets both the technical and scientific needs of the project
      \end{itemize}

    \item Training
    \begin{itemize}
        \item Use certification to document expertise and encourage learning new skills
        \item Encourage training in software and computing as a continuing physics activity
        \item Use mentors to spread scientific software development standards
        \item Involve computing professionals in the training of scientific domain experts
        \item Use online media to share training
        \item Use workbooks and wikis as evolving, interactive software documentation
        \item Provide young scientists with opportunities to learn computing and software skills that are marketable for non-academic jobs
    \end{itemize}

\end{itemize}


\section{Introduction}
\label{sec:Intro}

Over time, \HEP projects (experimental and theoretical)
have become larger and more complex, and generate ever larger datasets, and require ever higher
precision analysis.  In addition, the breadth of projects scientifically relevant to Particle Physics has increased to
include fields such as observational cosmology and deep underground experiments.
In order to deal with these changes, \HEP software has by necessity become more complex.
Software complexity has also increased due to evolving language standards, which provide more options to programmers.
Limitations on power consumption are changing the architecture of computer hardware,
requiring fundamental changes in software design to continue the ``Moore's law'' scaling of computing performance vs cost.
Increased software complexity is measured not just in
the number of lines of code, but in the sophistication of the algorithms employed, and in the diversity and
breadth of the software technologies employed.

The increasing sophistication of \HEP software has had demonstrable positive effects on the scientific output of our field
such as by improving the accuracy
of simulations, and by improving the efficiency and precision of scientific conclusions extracted from data using
techniques like Multi-Variate-Analysis.
Improved software organization, management, and testing standards have also benefited scientific output by allowing coherent and consistent data sets for analysis to be produced within hours of logging.

Increasingly complex software also has costs.  For instance, software development now takes a significant fraction
of the engineering and operations resources required to perform
\HEP scientific research.  Additionally, for scientists to
make effective use of complex software, more training, and more professional-level support are often required.
Increasingly complex software can also have direct financial costs, such as requiring more computing power to
execute, through commercial software license fees, and by requiring professionally-trained computer support personnel.
The trends of increasing software sophistication, evolving hardware platforms, and
increasing reliance on software for scientific progress show no signs of reversing in the near future.
The challenge for the next decade is then to guide the development of HEP software so as to maximize the scientific
output, while respecting the real resource constraints experienced by the field.

The primary responsibility for the scientific success of an HEP project lies with the scientists organizing
and collaborating on it.  This includes the responsibility for the major software design, organization,
and support decisions which every project must make.  However, no modern project can write its software completely
``from scratch''.  To the extent that projects have common features, sharing of common solutions can help reduce the
cost and risk of their software development.  Furthermore, the development cost of some common software is beyond the scope of
a single project, requiring a broader community of contributors to create and maintain the software that benefits
them all.

In discussions amongst ourselves and with members of the \HEP community,
these were the most common themes. Put more succinctly, the larger
goals for the next decade in the area
of software development and training are:

\begin{itemize}
\item To maximize the scientific productivity of our community
in an era of reduced resources, we must use
software development strategies and staffing models that will result in products
that are generally useful for the wider \HEP community.
\item We must respond to the evolving technology market, especially
with respect to computer processors, by
developing and evolving software that will perform with optimal efficiency
in future computing systems.
\item We must insure that our developers and users will have the
training needed to create, maintain, and use the increasingly complex software
environments and computing systems that will be part of future \HEP projects.
\end{itemize}

In the rest of this document we review the discussions in various 
areas related to software development and make specific a
number of specific recommendations to achieve these goals.





\section{Software Toolkits and Reuse}
\label{CpFI4:sec:reuse}

Software is a dynamic environment, where languages, infrastructure,
algorithms, and requirements are all evolving rapidly.
For scientific computing to be effective and competitive in this environment,
scientific software must therefore be continually renewed.
However, just as new projects often reuse or re-`purpose hardware
from previous projects, so too can software
developers reuse earlier software. 

 Software reuse has several potential benefits, such as:
\begin{itemize}
\item Reduced development time.  In particular, the lead time to having
functional software on a new project can often be greatly reduced.
\item Reduced cost to the project through lower demand on software developers.
\item Reduced risk through use of proven algorithms and implementations.
\item Preservation of detailed scientific knowledge expressed in the software implementation.
\end{itemize}
Software reuse also presents additional costs and risks, such as:
\begin{itemize}
\item Older code must often be refurbished to meet modern requirements,
which can mean migration, refactoring, or even a complete rewrite.
\item Software developers are often reluctant to reuse code,
as the effort required to understand poorly documented or
poorly-implemented code can be greater than the effort needed to write new code.
\item Commitment to code reuse can place an project at risk
if an external developer stops supporting the code, or develops the
code in a direction away from that project's needs.
\end{itemize}

Software reuse only makes sense if the benefits exceed the costs.
That decision will be made individually by each new project.
All current HEP projects rely upon some combination of new and reused
software. The choice of what software to reuse and which to develop
new is driven by both technical and non-technical considerations.
In an era of increasingly constrained budgets,
it makes sense to try to organize software development
in \HEP in a way that maximizes potential future
reuse across the field.  This would allow the community
to reduce cost and risk as a whole, over time.
We have therefore identified several models in which
\HEP software development could evolve in the future to maximize the chance
of future reuse, as described below.

One obvious case of successful reuse is toolkits.  Toolkits form around mature problems,
where the community has reached consensus on the major goals and methods,
and so is able to work coherently on the solution.
Toolkits are supported through the ``open source'' model, which amortizes
the cost across many projects (and sometimes fields).
This guarantees continual expert presence for maintenance and updating.
It also allows natural evolution, as the population
of developers represents the interests of current users.
The existing highly-successful toolkits (ROOT, Geant, RooFit, ...) nucleated at
at large institutions which provided steady initial support.
Several areas of \HEP software have the potential
for becoming toolkits, such as fast Monte Carlos, Monte Carlo generators, 
geometry description packages,
event-processing frameworks, databases, and track reconstruction algorithms (among others).
It is possible that a supportive institutional environment
could precipitate formation of a toolkit in some of these areas. 

\begin{itemize}
\item[] {\bf Recommendation:} 
Continue to support established toolkits (Geant4, ROOT, ...)
\item[] {\bf Recommendation:} 
Encourage the creation of new toolkits from existing successful common software
\end{itemize}

Another form of reuse is through software sharing.  Code is often copied from one project,
and modified to serve a new, related purpose on another.  While
this is not as coherent and doesn't offer as much long-term
potential benefit to the field as reuse through toolkit creation,
it can still offer big benefits to specific projects.
It is also more flexible than toolkit creation, as the copied code
is ``owned'' by the new project, which can modify it without any external constraints.

From discussions with our constituents, it became clear that successful
software sharing requires developer continuity; at least
initially, expert developers active in the previous project must be available
to adapt the copied software for the new project.
This works most successfully when the developers are scientific members
of both previous and new projects.  It becomes problematic
when the experts are computing professionals, supported by project funds.
The need to assign the experts efforts to one or
the other project's budget artificially compartmentalizes the problem,
when the benefit is to the community as a whole.
A possible way around this would be if computing professionals could be
matrixed across several projects while supporting the
transfer of software from one project to the other.

A related case arises in Monte Carlo generators, which
have become a computationally intensive component of the Monte
Carlo simulation workflow on the LHC experiments.  A wide range of event
generators are being developed in the US and elsewhere to meet the
changing needs of experimental work, and to capture the latest
theoretical developments for use by the experiments.   Because the
development teams are small (often only 1 or 2 people), they
often are not able to address all the technical computational
issues required for efficient performance using GRID
computing resources.  In experimental software, these technical issues
are often addressed by software professionals, who are integrated with the
scientific development community.
Allowing computing professionals nominally associated with the experimental program
to integrate with theorists involved in the generator development teams would be an effective
way to improve computational efficiency using existing resources.

\begin{itemize}
\item[] {\bf Recommendation:} 
 Allow flexible, reliable funding of software experts to facilitate transfer of software and sharing of technical expertise between projects
\end{itemize}

Projects typically restrict read access to their software to members, as a necessary
way of protecting the intellectual content of their work from competitors.
This however makes software sharing less likely.  In particular,
because non-members cannot search project code repositories,
they cannot discovery existing solutions to similar problems.
One way to improve this situation would be to encourage projects to put
non-sensitive code in read-accessible repositories.  Projects should also
use public licensing to protect their code and facilitate reuse.  Finally,
new projects should define a 'twilight' clause in their membership agreements,
whereby all the project's software will be made publicly read accessible
at some date after the last data has been recorded.

\begin{itemize}
\item[] {\bf Recommendation:} 
Facilitate code sharing through open-source licensing and use of publicly-readable repositories
\end{itemize}

Use of different software management tools can form a barrier to software sharing across projects.
Different tools can also adversely affect developers and users when they
transition between projects.
To the extent possible, use of common management tools
and common computational environment tools will reduce these burdens.

\begin{itemize}
\item[] {\bf Recommendation:} 
Consolidate and standardize software management tools to minimize cross-project ``friction''
\end{itemize}

\section{Software Development and Hardware Architecture}
\label{CpFI4:sec:development}

Over the past 15 years the bulk of software development in high energy physics
has shifted to programming on commodity x86 processors with Linux as the
main operating
system. The software development models of most large (and many small) projects
shifted to C++ and object oriented programming, with an emphasis on flexibility
for code evolution, maintainability and in some cases the requirements of allowing larger
numbers of distributed collaborators to contribute to the code base. Most high
energy physics applications are designed as simple sequential programs,
many instances of which can be run in embarrassingly parallel fashion, typically
referred to as high throughput computing (HTC). 

Exponential increases in luminosity and dataset sizes over the lifetime of a
given experiment and/or between successive generations of experiments
is also a critical ingredient which drives our science. These increases provide
the improved statistical power necessary to perform many measurements.
To perform such measurements without similar increases in the cost of
the computing, we have relied on the exponential increases in computing
power per unit cost (and similarly for storage, memory, etc.) famously
described by Gordon Moore (as Moore's law). Recent years however have seen a
significant change in the evolution of processor design relative to the
previous decades~\cite{GAMEOVER}. Realizing these exponential gains
in processor performance per unit cost will be much more difficult in
the future than over the past few decades.

In recent years, technology limitations, in particular
regarding power consumption, have triggered profound changes in the
evolution of computing processor technology. In the past software
could be run unchanged on successive processor generations and
achieve Moore's Law-like performance gains. This behavior has
allowed software designs based on simple, sequential programming
models to scale easily through enormous increases in
performance. The era of scaling for such sequential applications is
now over. The limitations on power consumption are leading to a new
era in which scalability will need to be achieved via significantly
more application parallelism and the exploitation of specialized
floating point capabilities. Achieving these huge potential
increases will transform completely the processor landscape and
software design. Failure to adapt will imply an end to the
exponential cost reductions for computing which have been
fundamental to enabling the progress of science in general and
specifically in high energy physics.

Previously one could expect to take a given code, and often the
same binary executable, and run it with greater computational
performance on newer generations of processors with roughly exponential
gains over time as described by Moore's Law.  A combination of increased
instruction level parallelism and (in particular) processor clock frequency
increases insured that expectations of such gains could be met in generation
after generation of processors. Over the past 10 years, however,
processors have begun to hit scaling limits, largely driven by overall
power consumption.

The first large change in commercial processor products as a
result of these limits was the introduction of ``multi-core'' CPUs,
with more than one functional processor on a chip. At the same time
clock frequencies ceased to increase with each processor generation and
indeed were often reduced relative to the peak. The result of this was
one could no longer expect that single, sequential applications would
run faster on newer processors. However in the first approximation,
the individual cores in the multi-core CPUs appeared more or less
like the single standalone processors used previously. Most large
scientific applications (HPC/parallel or high throughput) run in
any case on clusters and the additional cores are often simply
scheduled as if they were additional nodes in the cluster. This
allows overall throughput to continue to scale even if that of a
single application does not. It has several disadvantages, though,
in that a number of things that would have been roughly constant
over subsequent purchasing generations in a given cluster (with
a more or less fixed number of rack slots, say) now grow with each
generation of machines in the computer center. This includes the
total memory required in each box, the number of open files and/or
database connections, increasing number of independent (and incoherent)
I/O streams, the number of jobs handled by batch schedulers,
etc.  The specifics vary from application to application, but
potential difficulties in continually scaling these system parameters
puts some pressure on applications to make code changes in response,
for example by introducing thread-level parallelism where it did
not previously exist.

There is moreover a more general expectation that the limit of power
consumption on future Moore's Law scaling will lead to more profound
changes going forward. In particular, the power hungry x86-64
``large'' cores of today will likely be replaced wholly or in part by
simpler and less power hungry ``small'' cores. These smaller cores
effectively dial back some of the complexity added, at the expense of
increased power, in the period when industry was still making single
core performance scale with Moore's Law.  The result is expected to be
ever greater numbers of these smaller cores, perhaps with specialized
functions like large vector units, and typically with smaller memory
caches than the ``large'' cores.  Exploiting these devices fully will
also push applications to make larger structural code changes to
introduce significantly more fine-grained parallelism.

\begin{itemize}
\item[] {\bf Recommendation:} Significant investments in software
are needed to adapt to the evolution of computing processors, both
as basic R\&D into appropriate techniques and as re-engineering
``upgrades''
\item[] {\bf Recommendation:} New software should be designed, and 
existing software re-engineered, to expose parallelism at multiple levels
\end{itemize}

Although it is very hard to predict precisely where the market will wind up
in the long run, we already see several concrete examples which give
indications as to the kinds of things that we will see going forward, such
as Intel's MIC architecture, increased interest in low power ARM processors
for the server market, and General Purpose Graphics Processing Units 
(GPGPU's or GPU's). Overall the market is likely to see significantly more 
heterogeneity in products than in the past couple of decades. Effectively 
exploiting these newer architectures will require changes in the software
to exhibit significantly more parallelism at all levels, much improved locality
of access to data in memory and attention to maximize floating point
performance. Most of the scientific software and algorithms
in use today in projects was designed for the sequential processor
model in use for many decades and require significant re-engineering to 
exploit properly these new architectures.

Most physics algorithm-level software for projects is written by
physicists, rather than by software professionals. Adapting to this new
reality of hardware heterogeneity, and complex programming models will
be a challenge for many physicists who do not have the time nor
expertise to optimize their code to multiple hardware architectures

\begin{itemize}
\item[] {\bf Recommendation:} Develop flexible software architectures that can exploit efficiently a variety of possible future hardware options
\item[] {\bf Recommendation:} Support investigation and development of tools that allow user-level code to run on multiple, diverse hardware architectures
\end{itemize}

A broad and balanced mix of
effort on a number of elements will be required~\cite{ARCHWP}, including
general
investigations into newer processor architectures and programming
models, the simulation, pattern recognition algorithms in the
experiment trigger and reconstruction, tools and systems and
analysis techniques. Many aspects of the areas to investigate are
not unique to HEP, but as always the needs of our
scientific research program compel us to work at the leading edge of
progress in computing technology. As deviations from Moore's Law
cost scaling are already becoming visible, we expect that the
efforts will require concrete upgrades to projects already in the
next few years, as well as R\&D for the longer term eventually aimed at
efficient scalability of our applications through order of magnitude
increases in processor power over the next decade.

Specifically, investments will be needed in common toolkits like Geant4,
ROOT and event generators to make them scalable and efficient on the 
newer architectures. Indeed some efforts are already underway at 
FNAL~\cite{FNALWP} (Geant4 and ROOT) and at SLAC~\cite{SLACWP} (Geant4), 
for example. The experience from these efforts will likely guide the way 
for other projects. 

In addition existing project specific codes, for example for triggering
and event reconstruction software, will likely need to evolve. In
some cases, where the time scales and dataset size increases will be
large, they may need to be re-engineered and/or rewritten.

To summarize, the evolution of processor technology will have a major
impact on software development over the next decade. 

\section{Staffing}
\label{CpFI4:sec:staffing}

Two decades ago during the running of LEP, CLEO and the Run 1
Tevatron experiments the dominant programming language was Fortran.
Most professors in the community knew how to program well enough
in it, and it was easy to train new graduate students to use the
language and tools, like CERNLIB, of that time.  Staffing of software
efforts for the core software of the projects came from laboratories
like CERN, FNAL, and Cornell.  In addition to framework skeletons
and foundation libraries each project had a small team of people
that would integrate and debug the contributions to applications
from university personnel. In the early 1990s as the complexity of these
software applications grew problems caused by the deficiencies
 of the language, and the tool sets 
available for it, developed. 
New projects like BaBar and the 4 LHC experiments were
designing C++ applications from scratch with the help of hired
software engineering efforts. More attention was paid to the training
of postdocs and graduate students. However there were some who did
not see software development as their primary task.  These people
along with many of their professors were left behind by this
transition.  It was no longer the case that anyone could contribute
to software efforts. Software engineering skills and an ability to
contribute to large, multi-million line, projects was required. The
transition was lead by key people within the projects with many
years of domain expertise that made an effort to learn the new
programming model and language and spread that knowledge to their
peers. It should be noted that in some frontiers like the cosmic
and intensity frontiers the above transition is still ongoing.

We are now on the threshold of a new transition, from a serial
programming model to a concurrent model. We as a field should examine
the lessons learned from the previous transition by looking at the
efforts that were most successful and not repeating the mistakes.

In preparation for writing this report, software leaders and
representatives from many projects from the energy, intensity,
and cosmic frontiers were interviewed.  All gave the same advice, based 
on their experience, regardless of which stage in the above transition or 
adoption of software engineering techniques they are in. These are 
summarized in the following principles:

\begin{itemize}
\item[] {\bf Recommendation:} Recognize software efforts as sub-projects of the project.  
\end{itemize}

Many past and existing projects have Software and
Computing organizations, however much of the effort comes from
students or postdocs who were hired to work on detector
projects or to work on analysis at a University, and whose software
contributions are incidental. They may learn
software engineering because they are interested in it, and see it
as a marketable skill, but many do not.  As the technology trends
raise the bar on required skills to effectively contribute, software
can no longer be treated as something you do after you've built a
detector or before you process and analysis dataset. 
It is key to have people who can reliably estimate the level
of effort needed for the project over the lifetime of the
project; both to be able to explain that need to the collaboration,
and to international funding agencies.

In our surveys we have identified three types of people that
contribute to software efforts: domain experts who know very little
about software and computing, computing professionals, and most
importantly the bridge people who usually came from the field but
took a career path that allowed them to get staff positions as
computational physicists. Computing professional input into the
implementation of the frameworks and algorithms is very important.
However we know of no team of isolated computing professionals who
delivered software of lasting value.  Computing professionals need
to be integrated as part of the project team over the life
time of the project and they should be grouped with and lead by
the bridge people identified above. These people can effectively
communicate requirements and speak the same language as the computing
professionals when implementation problems (bugs) need to be solved.
They can also speak to the domain experts, encouraging them to write
pseudo-code for how they would solve the algorithmic problem, then
help the computing professional understand, and rewrite that code
in the most efficient way for the targeted programming model.  This
model will not be intuitive or natural, so it seems that forming
these teams is the only way to produce high quality implementations
that meet the technical and physics requirements of the
projects.

\begin{itemize}
\item[] {\bf Recommendation:} Integrate computing professionals as part of the project team over the life of the project
\item[] {\bf Recommendation:} Integrate software professionals with scientist developers to insure software meets both the technical and physics performance needs of the project
\end{itemize}

User support staffing is very important for projects.  It
accelerates the ability of new people to contribute.  The best model
we have found is to form a triage system for user support.  An
individual that is dedicated to user support is identified to field
questions from the collaboration.  Through time that individual
will learn enough to be able to handle most questions by him or
herself without distracting the expert programmers.  When needed
the user support person can redirect questions to the appropriate
expert group or individual.

Staffing for the future challenges imposed by the changing technologies
discussed in previous sections will require a change in the way we
do business.  This reality needs to be taken into account for any
future new or upgraded projects.

\section{Training}
\label{CpFI4:sec:training}

We give our thoughts on training  based on  discussions within the
sub-panel and with others. The thinking is based not only on
experience but also on the trends in the changing needs of the
field, the educational courses given to  physics students at
university, and also the opportunity to benefit to those needing
to eventually find jobs outside of the field. There is consensus
that activities and investment in computational skill development
and training of the high energy workforce has, and should continue
to have, a highly productive impact on the scientific outcomes. It
is also clear that having training that best matches the needs
across the field is a challenge.

We are facing: An increasingly dynamic, evolving and innovative
technology landscape; The need for  computing and software developments
to be more closely tied into the development of the physics codes
-  especially in the areas of concurrency and diversity of hardware
resources to be used;  And an increasingly transient workforce due
to the total cost of the program and  budgetary constraints.   We
are also persuaded of the value of exploring a range of training
activities and mechanisms - including university, on-job, mentorship,
commercial and online. We do not cover the traditional, static,
documentation as a means of training. We see utility in more focus
on real-world based exercises, embedded expertise-based joint work,
and the broader set of interactions available through virtual
presence.

The people doing computing in physics collaborations today span a
range of responsibilities and roles, each of which has particular,
and often specific, training needs.  We identify the following
categories:  Physics algorithm and method development and validation;
Scientific computing framework, workflow data management and
distributed system development; Advanced computing hardware and
software architecture and engineering; And production system
deployment, integration, and support.

We give a list of suggested training components in the rest of this
section.  Within each category thought is needed to decide how
tailored each training activity should be.

\begin{itemize}
\item[] {\bf Recommendation:} Use certification to document expertise and encourage learning new skills.
\end{itemize}

We see benefit in encouraging formal learning and measurement of
computing skills for physics undergraduate and graduate students.
One way of doing this is to provide means for certification in
software engineering, design and languages as an offered, encouraged,
and perhaps eventually required part of degrees in high energy
physics - just as for instrumentation and electronics.
 
Many universities have created, or are creating, computational
science and/or research courses, often multi-disciplinary and some
offered specifically as part of a domain science curriculum.
Existing programs known to the authors  include Indiana's PhD minor 
in Scientific Computing ~\cite{IUGRAD} and the Princeton
University graduate computational certificate ~\cite{PUGRAD}. 
Such programs exist at many universities today and should be connected 
as directly as possible into the needs of the particle and astro 
physics programs.

\begin{itemize}
\item[] {\bf Recommendation:} Encourage training in software and computing as a continuing physics activity
\end{itemize}

We (at BaBar, Belle, Fermi, LEP, LHC, RHIC, SDSS, SLAC, Tevatron
etc) have seen positive results from continuing training (which
includes making sure there are no barriers to contribution) including
both initial training of incoming software developers and regular
forums and materials to update and extend the training taken.

It is beneficial to plan and provide working environments to ensure
that such activities follow through the whole lifetime of a
project.

\begin{itemize}
\item[] {\bf Recommendation:} Use mentors to spread scientific software development standards
\end{itemize}

In recent years projects have taken specific steps to embed
trained software architects and engineers within the physics
development groups and activities. This ongoing expert guidance and
feedback results in better initial designs and, as importantly,
continued attention and eyes on the quality and process of changes
as they are made - many times quickly and under time pressure.

\begin{itemize}
\item[] {\bf Recommendation:} Involve computing professionals in the training of scientific domain experts
\end{itemize}

Giving computing professionals a role in training scientific domain
experts as well as continuous engagement in reviewing and guiding
the outcomes provides additional  attention to aspects like quality
and sustainability of the  products.  Separation of responsibilities
can provide for most combined benefit from individual skills and
expertise.  A representative example is for the domain experts to
write pseudo-code of the physics algorithms and analyses followed
by computing experts making it efficient for the targeted environment,
new hardware technologies etc.  Another example is targeted attention
to structure,  design and documentation of codes to improve their
robustness against change in functionality  as well as in the actual
people  working on them, throughout the lifetime of the product.

\begin{itemize}
\item[] {\bf Recommendation:} Use online media to share training
\end{itemize}

There has been an explosion in the availability, quality and ease
of use of web-based and streaming videos, collaborative tools, and
other multi-media technologies and content. These are already part
of the common, daily expectation of the emerging generation of
scientific and professional workforce. Examples that should be
investigated for principles, strengths and lessons learned are: The
TED program~\cite{TED}; The Kahn Academy~\cite{KHANACAD}; And current
University  online course offerings.

Initial investments that encourage a variety of approaches and
projects will help us find the most effective ways of using multi-media
in training activities.

The field can also invest in the concept  of ``networked training''
where coverage includes  a combination of developing simple data
analyses in tandem with training a set of contributors in mostly
an online setting. The global interest and engagement in the results
from the Higgs search, dark matter, dark energy and other cosmological
and particle searches can be leveraged to increase interest in such
programs.  Existing examples include Gaia~\cite{GAIA} and and
Chain-Reds~\cite{CHAINREDS}. Also LSST is currently considering
such a program. Using a body like ICFA to facilitate and oversee
such a program would demonstrate the importance of training to the
health of the field and ensure broad  opportunity and utility.

\begin{itemize}
\item[] {\bf Recommendation:} Use workbooks and wikis as evolving, interactive software documentation
\end{itemize}

Exercised-based workbooks and hands-on training workshops have shown
good return on investment in producing good quality and in the
ability to repurpose and extend  existing software codes:
\begin{itemize}
\item LHC Data analysis workshops have been a great success and are a model that can be copied and ``regularized''
\item CMS and ATLAS workbooks have resulted in new entrants being able to contribute quicker and more effectively.
\item Coding scrums and fests have shown the ability for the best experts to influence a broad set of other developers.
\item The simulation and theory communities have a long, successful history of tutorials and (summer) schools on various aspects of high performance computing ~\cite{HPCTraining}.
\item Technical writers can increase the quality of the communication of information through contributing to the organization, review and testing of material.
\end{itemize}

Best would be an integrated program in all computing plans of workbooks that address both workshop activities and individual learning.

The ``physics job pyramid'' means that many physics students and
post-docs will need to find jobs outside of the field at some point
in their careers. Computing and software are hot skills for the
marketplace and there has long been a recognition that particle
physics is a place where deep training and experience are gained.
We need to pay attention to the match between the tools, languages,
and technologies in demand outside of physics and those we adopt
and use in the field. We can do this by increasing our adoption of
software packages used in other scientific fields and the broader
society. The use of Hadoop and HDFS for data storage and access are
recent examples of such packages.   Lectures, surveys, and articles
covering technologies, trends and opportunities in the marketplace
help to keep the information current and personnel engaged.

\begin{itemize}
\item[] {\bf Recommendation:} Provide young scientists with opportunities to learn computing and software skills that are marketable for non-academic jobs
\end{itemize}

\ {Acknowledgements}
We wish to thank the many people in the HEP and Particle-astro
community who took the time to explain their views on these subjects,
and who contributed many of the ideas presented here.

\end{document}